\documentclass[aps,prb,twocolumn,notitlepage]{revtex4-2}

\usepackage{amssymb}
\usepackage{graphicx}
\usepackage{hyperref}

\begin{document}


\title{Single-photon detection with a Josephson junction coupled to a resonator}

\author{Dmitry S. Golubev$^1$}
\author{Evgeni V. Il'ichev$^{2,3}$}
\author{Leonid S. Kuzmin$^{4,5}$}
\affiliation{$^1$Pico group, QTF Centre of Excellence, Department of Applied Physics, Aalto University, P.O. Box 15100, FI-00076 Aalto, Finland}
\affiliation{$^2$Leibniz Institute of Photonic Technology, Albert-Einstein-Stra{\ss}e 9 (Beutenberg Campus), 07745 Jena,Germany}
\affiliation{$^3$Novosibirsk State Technical University, 20 Karl Marx Avenue, 630073 Novosibirsk, Russia}
\affiliation{$^4$Chalmers University of Technology, Gothenburg 41296, Sweden}
\affiliation{$^5$Nizhny Novgorod State Technical University, Nizhny Novgorod 603951, Russia}

\begin{abstract}
We use semiclassical formalism to optimize a microwave single photon detector based on switching events of a current biased Josephson junction coupled to a resonator. In order to detect very rare events, the average time between dark counts $\tau_{\rm dark}$ should be maximized taking into account that the switching time $\tau_{\rm sw}$ should be sufficiently small. We demonstrate that these times can be tuned in the wide range by changing the junction parameters, and the ratios $\tau_{\rm dark}/\tau_{\rm sw} \sim 10^9$ can be achieved. Therefore, a junction-resonator arrangement can be used for detecting extremely low photon fluxes, for instance for searching galactic axions.
\end{abstract}

\maketitle

\section{Introduction}

The implementation of scalable superconducting solid-state systems offers promising perspectives for microwave quantum engineering \cite{You,Wendin,Arute,Grajcar}.
In this context, the effective single photon detectors are required to properly manipulate weak microwave signals.
This need is becoming crucial, in particular, for quantum communication \cite{Thew,Pogor} and for search of axions \cite{Kuzmin,Braine}
to test the consequences of the standard model of particle physics.

Several types of microwave single photon detectors have been experimentally realized so far.
One type is based on semiconducting quantum dots, in which photon absorption
causes an electron jump from one dot to another \cite{Astafiev,Gustavsson,Khan}. The efficiency of
such detectors operated in quantum coherent regime has been theoretically analyzed in Ref. \cite{Vavilov}.
The detectors of the second type rely on superconducting qubits with level spacing close
to the photon energy \cite{Schuster,Johnson,Nakamura,Besse}.
Yet another detector type is based on a Josephson junction with strongly hysteretic current-voltage
characteristics. The operation principle of this detector is simple -- an absorbed photon switches
the current biased junction from the superconduting to the resistive state, which results in a dc voltage
signal. Some applications of this effect has been already demonstrated, see Ref. \cite{Chen,Oelsner1,Oelsner2}.
All types of single photon detectors mentioned above have narrow frequency band, which is necessary
for capturing very low energy microwave photons. At present, broad band detectors like, for example,
transition edge sensors or kinetic inductance detectors, are not sufficiently sensitive 
to resolve single photons in microwave frequency range.

\begin{figure}
\includegraphics[width=0.8\columnwidth]{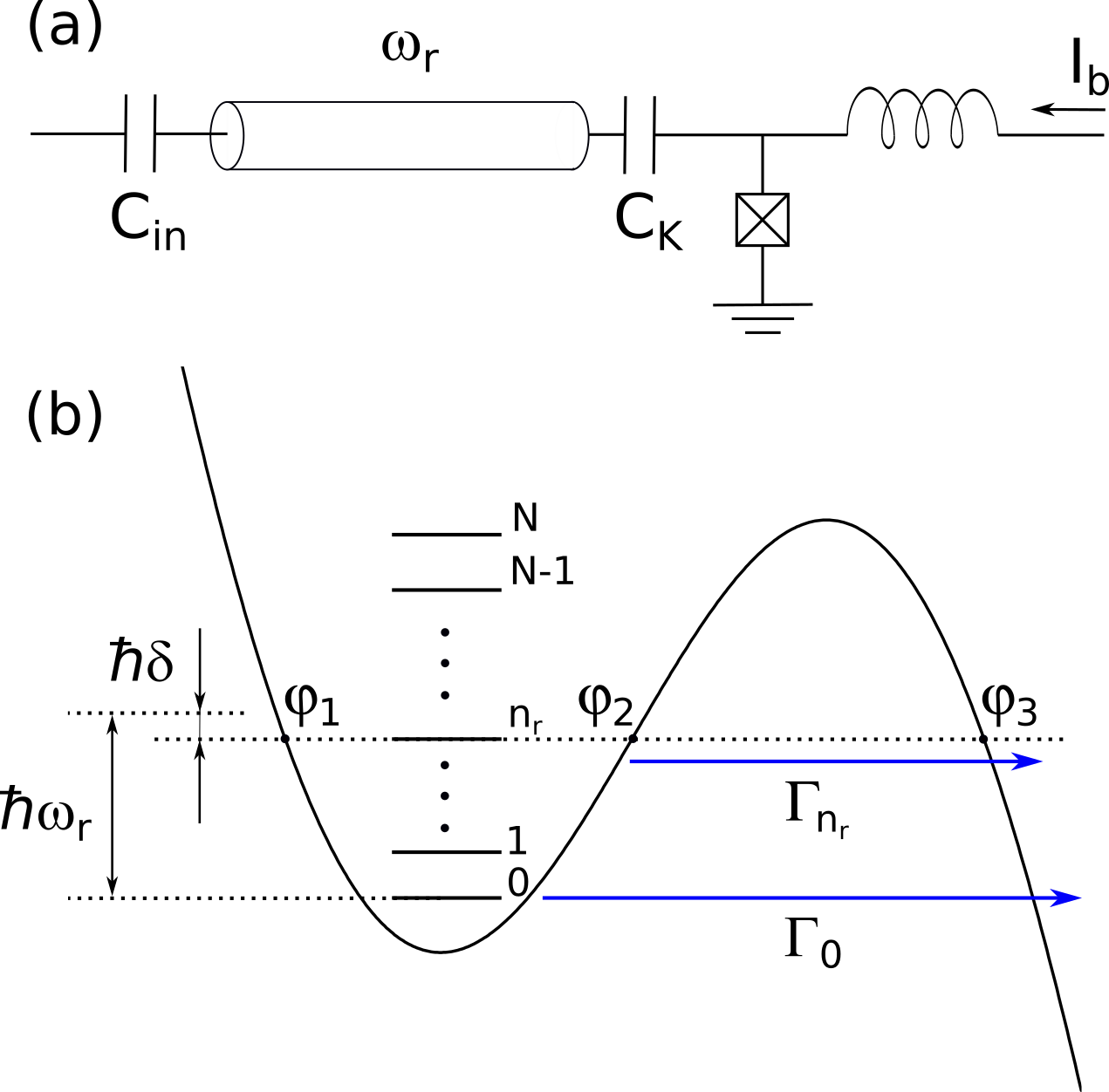}
\caption{(a) Josephson junction biased by the current $I_b$ and capacitively coupled to a resonator with the frequency $\omega_r$.
(b) Potential well of a tilted Josephson potential (\ref{U_phi}), which hosts $N+1$ descreet metastable energy levels
with the decay rates $\Gamma_n$, where $n=0,1,\dots N$.
The frequency of the resonator is close to the transition frequency between the levels 0 and $n_r$. }
\label{Fig1}
\end{figure}

Here we 
theoretically analyse a particular type of Josephson junction detector, which is supposed to operate at very low
photon fluxes and should wait for a photon arrival for a long time.
Accordingly, we require the detector to have the lowest possible dark count rate. At the same time,
the detector should very quickly switch to the resistive state after a detection event
in order to avoid photon loss and to enhance the detection efficiency. Such properties are
required for the detection of very rare events like, for example, decay of elementary particles.
A natural figure of merit for this type of detector is the ratio of the average time between the dark counts, $\tau_{\rm dark}$,
and the switching time $\tau_{\rm sw}$. For a good detector one should require $\tau_{\rm dark}/\tau_{\rm sw}\gg 1$.
We will demonstrate below that in a system with a junction coupled to a high quality factor resonator (see Fig. \ref{Fig1}(a))
one can achieve the ratios $\tau_{\rm dark}/\tau_{\rm sw} \sim 10^7$
with typical parameters of the setup provided the superconducting leads
of the junction, if made of aluminum, are cooled below $90$ \emph{mK}. In principle, one can even push this ratio to $10^9$.
In the latter case the junction having for example, the switching time
$\tau_{\rm sw}=1$ $\mu$s would have the dark count time $\tau_{\rm dark}=10^3$ \emph{s}.
We will show that while the times $\tau_{\rm sw}$ and $\tau_{\rm dark}$ can be tuned in the wide range by
changing the junction parameters, their ratio  predominantly depends on
the number of discreet energy levels in the Josephson potential well.
Theoretical model of a similar detector has been recently presented in Ref. \cite{Anghel},
where the minima of the two dimensional potential of the junction-plus-resonator system have been found,
the splitting between the energy levels in the potential wells have been determined and the dark count rate
of the detector has been roughly estimated as a switching rate of a weakly damped Josephson junction \cite{CL}. 
Here we extend the analysis of Ref. \cite{Anghel} in several ways. In particular, 
we include the transition matrix elements between the energy levels into the model
and solve the problem of the decay of metastable states localized in the potential wells in detail. 
In this way, we find not only the dark count rate of the detector, but also its' switching time.
We also analyze the effect of quasiparticles in the superconducting leads of the junction
and losses in the resonator on the detector performance.

Let us briefly discuss the operating principle of the detector.
Its schematics is presented in Fig. \ref{Fig1}(a).
The detector consists of a conventional $\lambda/2$ transmission line resonator coupled to a grounded Josephson junction via a capacitor $C_K$
and to an input circuit via a capacitor $C_{\rm in}$. An underdamped Josephson junction with strongly hysteric current-voltage characteristics is biased by the current $I_b$. This current can either be applied directly from the current source
or it can be induced by magnetic flux via a superconducting loop attached to the junction. 
The potential well of tilted Josephson potential hosts $N+1$ energy levels $E_n$, where the index $n$ changes from 0 to $N$.
By changing $I_b$ one can tune the system to a point where the condition
$\hbar\omega_r = E_{n_r}-E_0 + \hbar\delta$ is satisfied.
Here $\delta\ll \omega_r$ is a small detuning, which may differ from zero at the optimal operating point,
and $n_r$ is the number of the resonant level.
Once a photon is created in the resonator, the level $n_r$ gets populated after a short time
inversely proportional to the coupling strength between this level and the resonator, which we denote as $g_{0n_r}$.
In order to ensure fast switching, we require the decay rate of this level, $\Gamma_{n_r}$, to be large. This implies that the energy
$E_{n_r}$ should lie close to the top of the potential barrier. We find that preferably one should choose 
the second level from the top of the barrier with $n_r=N-1$ because at $n_r=N$ the dark count rate is enhanced due to
large escape rate of the highest level. On the one hand,
the total number of levels $N+1$ should be sufficiently large in order to make the dark count time
$\tau_{\rm dark}\propto \exp(36 N/5)$ as long as possible. On the other hand, $N$ cannot be too large in order to keep
the coupling $g_{0n_r}$ reasonably strong.

Below we present the theoretical model of the detector and discuss the trade-offs mentioned above in more detail.
Our analysis is based on the theory of inter-level transitions in the Josephson potential well,
which has been proposed by Larkin and Ovchinnikov \cite{LO1,LO2} and has been further developed in Refs. \cite{Kopietz,Chow,Zhang}.
We extend this theory by introducing the coupling between the junction and the resonator.
In experiment, the transitions between the energy levels of tilted Josephson potential have been detected
in Ref. \cite{Martinis1} by measuring the changes in the switching rate of the junction under microwave irradiation.
Later on, the setup of Fig. 1 was used to implement phase qubit, in which the two lowest energy levels in the well 
form the qubit and the higher levels are used for the readout \cite{Matrinis2,Martinis3}. 
Thus, the physics behind proposed photon detector is well established and tested in experiment.

The paper is organized as follows: in Sec. \ref{Model} we introduce the model, 
in Sec. \ref{Analytics} we provide an approximate analytical solution, 
in Sec. \ref{finite_T} we discuss how do finite temperature, dissipation in the resonator and
non-equilibrium quasiparticles limit the performance of the detector, 
in Sec. \ref{Numerics} we present the results of numerical simulation, and  in Sec. \ref{Summary}
we summarize the results. 

\section{Model}
\label{Model}

In this section we present the theoretical model of the system.
In Sec. \ref{Classical} we provide the classical equations of motion,
in Sec. \ref{Quantum} --- the corresponding quantum Hamiltonian,
in Sec. \ref{Semiclassics} we describe the semiclassical approximation,
and in Sec. \ref{Numerics1} we briefly describe the numerical procedure.
The derivation of the quantum Hamiltonian is presented in the Appendix \ref{App_H}.

\subsection{Classical equations of motion}
\label{Classical}

The classical dynamics of the system depicted in Fig. \ref{Fig1}(a) is described by the two coupled equations, which can be derived from
the Kirchhoff's laws,
\begin{eqnarray}
&& (C+C_{K})\frac{\hbar\ddot\varphi}{2e} + I_{qp}\left( \frac{\hbar\dot\varphi}{2} \right)
+ I_C\sin\varphi - C_{K}\dot V_{K} = I_b ,
\nonumber\\
&& \frac{2Z_0C_{K}\omega_r^3}{\pi} \frac{\hbar\dot\varphi}{2e} + \ddot V_{K} + \kappa_r\dot V_{K} + \omega_r^2 V_{K} = 0.
\label{sys}
\end{eqnarray}
Here $C$ is the junction capacitance, $I_{qp}(V)$
is the quasiparticle current through the junction,
$I_C= {\pi\Delta}/{2eR}$ is the Josephson critical current \cite{AB},
where $R$ is the normal state resistance of the junction and $\Delta$ is the superconducting gap,
$I_b$ is the bias current,
$C_K$ is the coupling capacitor between the junction and the resonator, $Z_0$ is the characteristic impedance of the 
transmission line resonator, 
and $V_K$ is the electric potential at the end of the resonator close to the capacitor $C_K$.
The frequency of the fundamental mode of the resonator
is $\omega_r=\pi/(t_0+Z_0C_K)$, where $t_0$ is the flight time of a photon through the resonator. 
The damping rate of the resonator $\kappa_r$ is composed of the internal losses
and the losses via the capacitor $C_{\rm in}$, $\kappa_r=\kappa_{\rm int}+(2/\pi) \omega_r^3 Z_0^2 C_{\rm in}^2$.
Eqs. (\ref{sys}) are valid provided $\omega_r Z_0C_K\ll\pi$ and
the quality factor of the resonator is high, $Q_r=\omega_r/\kappa_r \gg 1$.

\subsection{Quantum Hamiltonian}
\label{Quantum}

In this and in the next section we ignore the dissipation in the resonator setting $\kappa_r=0$, consider zero temperature limit, $T=0$,
and ignore the effect of quasiparticles.
These effects will be separately discussed in Sec. \ref{finite_T}.
The system of Fig. \ref{Fig1}(a) can be described by the quantum Hamiltonian (see Appendix \ref{App_H} for the derivation)
\begin{eqnarray}
\hat H = \hat H_J + \hat H_r + \hat H_{\rm int},
\label{H}
\end{eqnarray}
where
\begin{eqnarray}
\hat H_J = -4E_C\frac{\partial^2}{\partial\varphi^2} + U(\varphi),
\label{HJ}
\end{eqnarray}
is the Hamiltonian of the junction, $E_C=e^2/2(C+C_K)$ is the charging energy,
\begin{eqnarray}
U(\varphi) = - E_J\cos\varphi - \frac{\hbar I_b}{2e}\varphi + \frac{\hbar g^2}{\omega_r}\sqrt{\frac{E_J}{2E_C}}\varphi^2
\label{U_phi}
\end{eqnarray}
is the Josephson potential tilted by the bias current $I_b$ with
the quadratic correction coming from the coupling to the resonator, $E_J=\hbar I_C/2e$ is the Josephson energy,
\begin{eqnarray}
\hat H_r = \hbar\omega_r\left(\hat a^\dagger \hat a + \frac{1}{2}\right)
\label{H_r}
\end{eqnarray}
is the Hamiltonian of the resonator, and
\begin{eqnarray}
\hat H_{\rm int} = -i\hbar g\left(\frac{E_J}{2E_C}\right)^{1/4}(\hat a^\dagger - \hat a)\varphi
\label{Hint}
\end{eqnarray}
is the interaction term. Here
\begin{eqnarray}
g=\frac{C_K\omega_r^2}{\pi}\sqrt{\frac{Z_0 R_q}{2}}\left(\frac{E_C}{8E_J}\right)^{1/4}
\label{g}
\end{eqnarray}
is the coupling strength  between the resonator and the transition between
the levels 0 and 1 in the potential well of an unbiased junction, i.e. at $I_b=0$, and $R_q=h/e^2$ is the resistance quantum.
Although the expression (\ref{g}) is specific for the coplanar resonator capacitively
coupled to the junction, the Hamiltonian (\ref{H}-\ref{Hint}) is quite general and describes various types 
of resonators and couplings. The specifics of a particular setup affects only the expression for the coupling
constant $g$. In addition, for certain types of coupling the combination $-i(\hat a^\dagger - \hat a)$
in the interaction term (\ref{Hint}) should be replaced by the sum $\hat a^\dagger + \hat a$.

If one retains only the two lowest levels in a Josephson potential well with the energies $E_0$ and $E_1$
and considers zero bias current, $I_b=0$, the phase operator can be approximated as $\varphi \to (2E_C/E_J)^{1/4}\sigma_x$. Afterwards,
the Hamiltonian (\ref{H}) reduces to the usual Rabi Hamiltonian describing a transmon qubit \cite{Koch},
\begin{eqnarray}
\hat H_{\rm Rabi} &=& -\frac{E_1-E_0}{2}\sigma_z + \hbar\omega_r\left(\hat a^\dagger \hat a + \frac{1}{2}\right)
\nonumber\\ &&
-\, i\hbar g(\hat a^\dagger - \hat a)\sigma_x.
\label{H_Rabi}
\end{eqnarray}

\subsection{Semiclassical approximation}
\label{Semiclassics}

In the limit $E_J\gg E_C$ the energy levels $E_n$ can be obtained from Bohr-Sommerfeld quantization rule
\begin{eqnarray}
\int_{\varphi_1}^{\varphi_2} d\varphi \sqrt{\frac{E_n-U(\varphi)}{4E_C}} = \pi\left(n+\frac{1}{2}\right).
\label{En}
\end{eqnarray}
Here $\varphi_{1,2}$ are the classical turning points such that $U(\varphi_{1,2})=E_n$   (see Fig. \ref{Fig1}(b)).

At finite bias current the levels become metastable. The corresponding decay rates
can also be found from the semiclassical approximate expression,
\begin{eqnarray}
\Gamma_n = \frac{\omega(E_n)}{2\pi}\exp\left[ -2\int_{\varphi_2}^{\varphi_3} d\varphi \sqrt{\frac{U(\varphi)-E_n}{4E_C}} \right].
\label{Gamma}
\end{eqnarray}
Here $\omega(E_n)$ is the classical oscillation frequency in the potential well, which depends on the level energy $E_n$,
\begin{eqnarray}
\omega(E_n)=\frac{4\pi}{\hbar}\left[\int_{\varphi_1}^{\varphi_2} \frac{d\varphi}{\sqrt{E_C[E_n-U(\varphi)]}}\right]^{-1},
\end{eqnarray}
and $\varphi_3$ is the turning ponit on the other side of the potential barrier (see again Fig. \ref{Fig1}(b)).

For values of the bias current close the crictical one, $I_C-I_b\ll I_C$,
one can approximately replace the potential (\ref{U_phi}) by the cubic polynomial,
\begin{eqnarray}
U=\frac{E_J}{2}\left[ \alpha\phi^2 - \frac{\phi^3}{3} \right],
\label{U3}
\end{eqnarray}
where $\phi=\varphi-\varphi_{\min}$ is the deviation of the phase from its equilibrium value $\varphi_{\min}$ at which
the potential $U(\varphi)$ has the minimum, and
\begin{eqnarray}
\alpha=\sqrt{\frac{4\pi\hbar g^2}{\omega_r\sqrt{8E_JE_C}}+1-\frac{I_b^2}{I_C^2}}.
\end{eqnarray}

We also need to know the matrix elements $\varphi_{mn}$ of the Josephson phase evaluated between the wave functions of the levels $m$ and $n$.
We again  use the semiclassical approximation, which leads to the following expression:
\begin{eqnarray}
\varphi_{mn} = \frac{\omega(E_{mn}^+)}{2\pi}\int_{-\pi/\omega(E_{mn}^+)}^{\pi/\omega(E_{mn}^+)}
dt\, \varphi_{\rm cl}(t,E_{mn}^+)
\nonumber\\ \times\,
\sin\big[ (m-n)\omega(E_{mn}^+)t \big].
\label{integral}
\end{eqnarray}
Here $E_{mn}^+=(E_m+E_n)/2$ and $\varphi_{\rm cl}$ is the solution of the classical equation of motion
\begin{eqnarray}
C\frac{\hbar\ddot\varphi_{\rm cl}}{2e} + I_C\sin\varphi_{\rm cl}
+ \frac{2\hbar g^2}{\omega_r}\sqrt{\frac{E_J}{2E_C}}\varphi_{\rm cl} = I_b
\end{eqnarray}
with the energy $E_{mn}^+$. The intergral (\ref{integral}) can be solved analytically \cite{LO1,LO2} for  $I_C-I_b\ll I_C$, where
the cubic approximation (\ref{U3}) holds. In this case one finds
\begin{eqnarray}
\varphi_{mn} = \frac{\pi^2\sqrt{3}}{2}\left(\frac{12 U_b}{E_J}\right)^{1/3}
\frac{\cos\left(\frac{\pi}{6}-\theta\right)}{K^2(k)}
\nonumber\\ \times
\frac{(-1)^{|m-n|+1}(m-n)}{\sinh\left[ \pi(m-n)\frac{K\left(\sqrt{1-k^2}\right)}{K(k)} \right]}.
\label{phi_mn}
\end{eqnarray}
Here  $U_b=2E_J\alpha^{3/2}/3$ is the height of the potential barrier, $K(k)$ is the complete elliptic integral,
\begin{eqnarray}
k=\sqrt{\frac{\sin\theta}{\cos\left(\frac{\pi}{6}-\theta\right)}}
\end{eqnarray}
is the ellipticity parameter, and
\begin{eqnarray}
\theta = \frac{2}{3}\arcsin\sqrt{\frac{E_{mn}^+-U(\varphi_{\min})}{U_b}}.
\end{eqnarray}

The semicalssical expressions (\ref{Gamma},\ref{phi_mn}) become invalid
close to the top of the potential barrier,  where one should use other approximations \cite{LO2,Kopietz}.
Here we will avoid doing that in order to keep the model simple.
Moreover, as we have mentioned above,
the optimal choice for the resonant level is $n_r=N-1$.
This level lies sufficiently far from the barrier top, where the approximation (\ref{phi_mn}) is still applicable.
For the same reason,  we will ignore the transition matrix elements between
the discreet levels and the continuous spectrum above the barrier top.

\subsection{Numerical solution of the problem}
\label{Numerics1}

Ideally, the dark count rate and the rate of switching should be found by solving the time evolution equation for the
wave function of the system,
\begin{eqnarray}
i\hbar\frac{\partial\Psi_k(t,\varphi)}{\partial t} = \sum_{p=0}^\infty \hat H_{kp}(\varphi)\Psi_p(t,\varphi),
\label{Schrodinger}
\end{eqnarray}
where the indexes $k$ and $p$ enumerate the number of photons in the resonator and $ \hat H_{kp}(\varphi)$ is the sub-block
of the Hamiltonian $\hat H$ relating the states with $k$ and $p$ photons. The initial wave function should be chosen such that
the junction intially finds itself in the ground state. The resonator should intially be in the ground state with zero photons
if one is interested in the dark count rate, or it should host one photon if one finds
the switching rate. Having solved Eq. (\ref{Schrodinger}),
one should evaluate the integral
\begin{eqnarray}
P(t) = \sum_{k=0}^\infty\int_{\varphi_1}^{\varphi_2} |\Psi_k(t,\varphi)|^2,
\label{P_exact}
\end{eqnarray}
which determines the probability for the Josephson phase to stay between the classical turning points $\varphi_1$ and $\varphi_2$
at time $t$ or, in other words,  the probability for the switching event not to occur before the time $t$.
Since the states in the potential well are metastable, the function (\ref{P_exact}) decays.
Depending on the choice of the initial conditions, the time scale of this decay defines either
the average time between the dark counts, $\tau_{\rm dark}$, or the switching time $\tau_{\rm sw}$.

Since Eq. (\ref{Schrodinger}) is difficult to solve, we make the usual set of approximations. 
First, we expand the wave function of the system in the basis of the wave functions $\psi_n(\varphi)$
of the levels in the Josephson potential well,
\begin{eqnarray}
\Psi_k(t,\varphi) = \sum_{n=0}^{N} c_{nk}(t) \psi_n(\varphi).
\end{eqnarray}
Afterwards, we add the imaginary part $-i\Gamma_n/2$ to the energy of each discreet level, $E_n \to E_n-i\Gamma_n/2$.
The Hamiltonian $\hat H$ then
acquires the form of a non-Hermitian matrix describing the decay of the initial metastable state in time.
We  assume that the coupling constant (\ref{g}) is small,
$g\ll \omega_r\left( 2E_C/E_J \right)^{1/4}$.
This allows us to restrict the number of photons in the resonator by  three values $k=$0, 1, 2.
Defining the coupling constants between the resonator and the interlevel transitions $m\leftrightarrow n$ as
\begin{eqnarray}
g_{mn}=g(E_J/2E_C)^{1/4}\varphi_{mn},
\label{g_mn}
\end{eqnarray}
we write the Hamiltoninan (\ref{H}) in the form
of a non-Hermitian $3(N+1)\times 3(N+1)$ matrix,
\begin{widetext}
\begin{eqnarray}
\tilde H = \left(\begin{array}{ccc}
\left(E_n+\frac{\hbar\omega_r}{2}-i\frac{\hbar\Gamma_n}{2}\right)\delta_{mn} & i\hbar g_{mn} & 0 \\
-i\hbar g_{mn} & \left(E_n+\frac{3\hbar\omega_r}{2}-i\frac{\hbar\Gamma_n}{2}\right)\delta_{mn}  & i\sqrt{2}\,\hbar g_{mn} \\
0 & -i\sqrt{2}\,\hbar g_{mn} & \left(E_n+\frac{5\hbar\omega_r}{2}-i\frac{\hbar\Gamma_n}{2}\right)\delta_{mn}
\end{array}\right).
\label{Happ}
\end{eqnarray}
\end{widetext}
The $(N+1)\times (N+1)$ sub-blocks on the diagonal of this matrix contain the complex energies  
of the metastable levels $E_n-i\Gamma_n/2$.
Namely, the sub-block in the top left corner of the matrix describes the state with zero photons in the resonator, 
the sub-block in the middle --- the state with two photons,
and the sub-block in the right bottom corner --- the state with three photons.
The off-diagonal sub-blocks originate from the interaction Hamiltonian (\ref{Hint}) and contain the matrix
elements (\ref{g_mn}). In this approximation,
the probability (\ref{P_exact}) for the system to stay in the initial state $\Psi_0$ becomes
\begin{eqnarray}
P(t) = \left| e^{-\tilde Ht}\Psi_0 \right|^2.
\label{Pt}
\end{eqnarray}
In order to find the dark count rate, we choose $\Psi_0$ in the form of $3(N+1)$-dimensional vector
with all matrix elements equal to zero except for the first one, $\Psi_0^{T}=(1,0,\dots 0)$.
The switching rate should be determined by setting all the components of the initial wave function 
$\Psi_0$ to be zero except the component with the number $N+2$, which should be equal to 1. By our convention, this component
corresponds to the ground state of the junction and one photon in the resonator.

\section{Approximate analytical results}
\label{Analytics}

Before proceeding to the numerical evaluation of the dark count and switching rates,
we present simple analytical approximations, which may be useful for optimizing the detector parameters.
We first consider the dark count rate. In the lowest order of the perturbation theory in the coupling strength $g$,
the ground state wave function is the product of the state with zero photons  in the resonator, $k=0$, and of the ground state in
the Josephson potential well with $n=0$. We denote this state as $|00\rangle$.
Applying second order perturbation theory in $g$ to the Hamiltonian $\tilde H$ (\ref{Happ}),
we find the corrected energy of this state in the form
\begin{eqnarray}
\tilde E_{00} = E_0 - i\frac{\hbar\Gamma_0}{2}
+ \sum_{s=0}^N \frac{\hbar^2 g_{0s}^2}{E_0 - E_s-\hbar\omega_r + i\frac{\hbar(\Gamma_s+\Gamma_0)}{2}}.
\label{E00}
\end{eqnarray}
Taking the imaginary part of this expression, we find the dark count rate  of the device at zero temperature
$\Gamma_{\rm dark} =\tau_{\rm dark}^{-1} = -({2}/{\hbar})\,{\rm Im}\, \tilde E_{00}$, 
\begin{eqnarray}
\Gamma_{\rm dark}  = \Gamma_0
+ \sum_{s=0}^N \frac{g_{0s}^2(\Gamma_s+\Gamma_0)}{\left( \frac{E_s-E_0}{\hbar}+\omega_r\right)^2 + \frac{(\Gamma_s+\Gamma_0)^2}{4}}.
\label{Gamma_dark}
\end{eqnarray}
The second term in this expression describes the increase of the dark count rate due to the admixture of the excited states
in the potential well caused by the interaction between the junction and the resonator.
The approximate expression (\ref{Gamma_dark}) can be used if $g\lesssim E_1-E_0$.

In order to find the switching rate
we consider the state with one photon in the resonator ($k=1$) and the ground state of the junction ($n=0$), i.e. the state $|10\rangle$,
and the state with zero photons in the resonator ($k=0$) and the junction excited to the level $n_r$, i.e. the state $|0n_r\rangle$.
These two states are almost degenerate. We separate the two dimensional  sub-space
spanned by them and approximately write the Hamiltonian (\ref{Happ}) as a $2\times 2$ matrix
\begin{eqnarray}
\tilde H' = \left( \begin{array}{cc} \tilde E_{0n_r} + \frac{\hbar\omega_r}{2} - i\frac{\hbar \tilde\Gamma_{0n_r}}{2} & -ig_{0n_r} \\
ig_{0n_r} & \tilde E_{10} + \frac{3\hbar\omega_r}{2} - i\frac{\hbar\tilde\Gamma_{10}}{2}  \end{array} \right).
\label{H_app}
\end{eqnarray}
Here the corrected energies and the decay rates are
\begin{eqnarray}
\tilde E_{0n_r} = E_{n_r}
+ \sum_{s=1}^N\frac{\hbar^2 g_{sn_r}^2(E_{n_r} - E_s-\hbar\omega_r)}
{(E_{n_r} - E_s-\hbar\omega_r)^2 + \frac{\hbar^2(\Gamma_s+\Gamma_{n_r})^2}{4}},
\label{E_nr}
\end{eqnarray}
\begin{eqnarray}
\tilde \Gamma_{0n_r} = \Gamma_{n_r}
+ \sum_{s=1}^N\frac{\hbar^2 g_{sn_r}^2(\Gamma_s+\Gamma_{n_r})}
{(E_{n_r} - E_s-\hbar\omega_r)^2 + \frac{\hbar^2(\Gamma_s+\Gamma_{n_r})^2}{4}},
\label{Gamma_nr}
\end{eqnarray}
\begin{eqnarray}
\tilde E_{10}  &=& E_{0}
+ \sum_{s\not=n_r}\frac{\hbar^2 g_{0s}^2(E_{0} + \hbar\omega_r - E_s)}{(E_{0} + \hbar\omega_r - E_s)^2 + \frac{\hbar^2(\Gamma_s+\Gamma_0)^2}{4}}
\nonumber\\ &&
+ \sum_{s=0}^N\frac{2\hbar^2 g_{0s}^2(E_{0}  - E_s - \hbar\omega_r)}{(E_{0}  - E_s - \hbar\omega_r)^2 + \frac{\hbar^2(\Gamma_s+\Gamma_0)^2}{4}},
\label{E_0}
\end{eqnarray}
\begin{eqnarray}
\tilde \Gamma_{10}  &=& \Gamma_{0}
+ \sum_{s\not=n_r}\frac{\hbar^2 g_{0s}^2(\Gamma_s+\Gamma_0)}{(E_{0} + \hbar\omega_r - E_s)^2 + \frac{\hbar^2(\Gamma_s+\Gamma_0)^2}{4}}
\nonumber\\ &&
+ \sum_{s=0}^N\frac{2\hbar^2 g_{0s}^2(\Gamma_s+\Gamma_0)}{(E_{0}  - E_s - \hbar\omega_r)^2 + \frac{\hbar^2(\Gamma_s+\Gamma_0)^2}{4}}.
\label{Gamma_0}
\end{eqnarray}
Note that the renormalized energies and the decay rates depend on the number of photons in the resonator. 

After approximate reduction of the Hilbert space to two dimensions,
the initial wave function $|10\rangle$ takes the form
$\Psi_0^T = (0,1)$, and the probability (\ref{Pt}) becomes
\begin{eqnarray}
&& P(t) \approx \left|e^{-i\tilde H't}\left(\begin{array}{c} 0 \\ 1 \end{array}\right)\right|^2
\nonumber\\ &&
=\,  e^{-\frac{(\tilde\Gamma_{0n_r}+\tilde\Gamma_{10})t}{2}} \left|\cos\Omega t
+ i\frac{\sqrt{\Omega^2-g_{0n_r}^2}}{\Omega} \sin\Omega t\right|^2.
\label{Pt2}
\end{eqnarray}
Here we have introduced the complex valued "frequency"
\begin{eqnarray}
\Omega = \sqrt{\left(\frac{\delta\omega}{2}+i\frac{\tilde\Gamma_{0n_r}-\tilde\Gamma_{10}}{4}\right)^2+g_{0n_r}^2},
\label{Omega}
\end{eqnarray}
where the detuning $\delta\omega$ is given by
\begin{eqnarray}
\hbar\delta\omega = \hbar\omega_r - \tilde E_{0nr}+\tilde E_{10}.
\end{eqnarray}
It differs from the detuning $\delta$, defined in Fig. 1(b), since the energies $\tilde E_{kn}$ are shifted
relative to the bare energies $E_n$.
The switching rate of the junction is given by the slowest decay rate of the function (\ref{Pt2}),
\begin{eqnarray}
\Gamma_{\rm sw} = \tau_{\rm sw}^{-1} = \frac{\tilde\Gamma_{0n_r}+\tilde\Gamma_{10}}{2}-2\,\big|{\rm Im}\,(\Omega)\big|.
\label{Gamma_sw}
\end{eqnarray}
The approximate expression (\ref{Gamma_sw})
is valid at sufficiently small detuning from the resonance, $\delta\omega\lesssim \omega_r/n_r$,
and for sufficiently weak coupling, $g\lesssim E_1-E_0$.

The switching rate (\ref{Gamma_sw}) reaches its maximum value at resonance  $\delta\omega=0$,
where it becomes
\begin{eqnarray}
\Gamma_{\rm sw}^{\max} = \frac{\tilde\Gamma_{0n_r}+\tilde\Gamma_{10}}{2}
-2{\rm Re}\left[\sqrt{\frac{(\tilde\Gamma_{0n_r}-\tilde\Gamma_{10})^2}{16}-g^2_{0n_r}}\right].
\label{Gamma_max}
\end{eqnarray}
Since the ground state level is much more stable than the level $n_r$, one can usually omit
the decay rate $\tilde\Gamma_{10}$ from Eq. (\ref{Gamma_max}). The maximum switching rate
is limited by the slowest bottleneck process. Indeed, at strong coupling $|g_{0n_r}|>\tilde\Gamma_{0n_r}/4$
it is limited by the decay of the $n_r$-th energy level, $\Gamma_{\rm sw}^{\max} = \tilde\Gamma_{0n_r}/2$, 
while in the weak coupling limit, $|g_{0n_r}|\ll \tilde\Gamma_{0n_r}/4$,
it becomes $\Gamma_{\rm sw}^{\max} = 4g_{0n_r}^2/\tilde\Gamma_{0n_r}$.

The dependence  of the switching rate on the detuning
(\ref{Gamma_sw}) has the form of a peak
of a rather unusual shape. It crosses over from the Lorentzian peak at weak coupling, $|g_{0n_r}|\lesssim \tilde\Gamma_{0n_r}/4$, to
the peak with a sharp cusp in the strong coupling regime $|g_{0n_r}|\gtrsim \tilde\Gamma_{0n_r}/4$, see Fig. \ref{plots}(b).
The half-width of this peak can be found exactly,
\begin{eqnarray}
\delta\omega_{1/2} &=& \sqrt{  \frac{16}{3} g^2_{0n_r} - \frac{\tilde\Gamma_{0n_r}^2}{4} },\;\;\; |g_{0n_r}|>\frac{\tilde\Gamma_{0n_r}}{4},
\nonumber\\
\delta\omega_{1/2} &=& 2F\sqrt{\frac{F^2 - \frac{g_{0n_r}^2}{2}}{F^2 + \frac{g_{0n_r}^2}{2}}},
\;\;\; |g_{0n_r}|<\frac{\tilde\Gamma_{0n_r}}{4}.
\label{D12}
\end{eqnarray}
Here we have  assumed that $\tilde\Gamma_{10}\ll\tilde\Gamma_{0n_r}$, and
introduced the combination
\begin{eqnarray}
F=\frac{\tilde\Gamma_{0n_r}}{4} + \sqrt{ \frac{\tilde\Gamma_{0n_r}^2}{16}  - g^2_{0n_r} }.
\end{eqnarray}
At very weak coupling $|g_{0n_r}|\ll \tilde\Gamma_{0n_r}/4$ we find $\delta\omega_{1/2}=\tilde\Gamma_{0n_r}$,
while in the opposite limit $|g_{0n_r}|\gg \tilde\Gamma_{0n_r}/4$ we obtain $\delta\omega_{1/2}=2|g_{0n_r}|/\sqrt{3}$.

Since the detuning $\delta\omega$ in our device is controlled by the bias current,
it makes sense to convert the half-width (\ref{D12}) into current units,
\begin{eqnarray}
\Delta I_{b,1/2} &=& \left|\frac{\hbar\delta\omega_{1/2}}{\frac{\partial (\tilde E_{n_r}-\tilde E_0)}{\partial I_b}}\right|
\nonumber\\
&\approx & \sqrt{\frac{2E_J}{E_C}}\frac{I_C}{I_{b,r}}
\left(1-\frac{I_{b,r}^2}{I_C^2}\right)^{3/4} \frac{e\delta\omega_{1/2}}{n_r}.
\label{I12}
\end{eqnarray}
Here $I_{b,r}$ is the value of the bias current, at which the resonance condition $\tilde E_{n_r}-\tilde E_0=\hbar\omega_r$
is achieved.

Let us now estimate the maximum ratio $\tau_{\rm dark}/\tau_{\rm sw}$. Assuming that the coupling is sufficiently strong,
\begin{eqnarray}
|g_{0n_r}|>{\tilde\Gamma_{0n_r}}/{4},
\label{condition}
\end{eqnarray}
and the junction is tuned to resonance, we obtain
\begin{eqnarray}
\frac{\tau_{\rm dark}}{\tau_{\rm sw}}\bigg|_{\max} = \frac{\Gamma_{\rm sw}^{\max}}{\Gamma_{\rm dark}}
= \frac{\tilde\Gamma_{0n_r}}{2\Gamma_{\rm dark}}.
\end{eqnarray}
We can roughly estimate the rates $\Gamma_{\rm dark}$ (\ref{Gamma_dark}) and $\tilde\Gamma_{0n_r}$  (\ref{Gamma_nr}) as
\begin{eqnarray}
\Gamma_{\rm dark} &\approx& \Gamma_0 + \frac{g_{0N}^2\Gamma_N}{(N+1)^2\omega_r^2},
\nonumber\\
\tilde\Gamma_{0n_r} &\approx& \Gamma_{n_r} + \frac{g^2_{Nn_r}\Gamma_N}{(N+1-n_r)^2\omega_r^2}.
\end{eqnarray}
Since we have assumed the condition (\ref{condition}) to be satisfied, the second terms in these experssions tend to dominate. Hence,
we obtain the following estimate valid for $n_r<N$,
\begin{eqnarray}
&& \frac{\tau_{\rm dark}}{\tau_{\rm sw}}\bigg|_{\max} \sim \frac{g^2_{Nn_r}}{2g_{0N}^2}
= \frac{\varphi^2_{Nn_r}}{2\varphi_{0N}^2}
\nonumber\\ &&
\approx  \frac{1}{2} \frac{(N-n_r)^2}{N^2}
e^{2\left(\pi + 2c_0  - \frac{c_0(1+2n_r)}{N} \right)n_r}.
\label{ratio1}
\end{eqnarray}
Here we have introduced the constant
\begin{eqnarray}
c_0 = \frac{\pi}{2\sqrt{3}}\left(2\frac{E\left({1}/{\sqrt{2}}\right)}{K\left({1}/{\sqrt{2}}\right)}-1\right)=0.4144\dots
\end{eqnarray}
For a given number of levels in the well, the ratio
(\ref{ratio1}) reaches the maximum value if one chooses  $n_r=N-1$, i.e. if one brings second closest to the barrier top
energy level in resonance with the resonator. In this case
\begin{eqnarray}
\frac{\tau_{\rm dark}}{\tau_{\rm sw}}\bigg|_{\max}
\sim   \frac{\exp\left[2\left(\pi +  \frac{c_0}{N} \right)(N-1)\right]}{2N^2}.
\label{ratio}
\end{eqnarray}
One can vary the absolute values of the times $\tau_{\rm dark}$ and $\tau_{\rm sw}$ by many orders of magnitude
by changing the critcal current, bias current or the charging energy of the junction. 
However, as Eq. (\ref{ratio}) shows, the ratio $\tau_{\rm dark}/\tau_{\rm sw}$
predominantly depends on the number of the levels in the potential well at resonance condition
irrespective of the specific values of $I_C,E_C$ or $I_b$.

\section{Effect of quasiparticles and damping in the resosnator}
\label{finite_T}

The dark count rate and the efficiency of the detector depend on the temperature of the
resonator environment, $T_r$, and on the temperature of the superconducting leads of the Josephson junction, $T_S$.
First, we consider the latter effect. At finite $T_S$ the quasiparticles present in the leads
cause up and down transitions between the neighboring energy levels of the Josephson potential well 
with the rates $\Gamma^{\rm qp}_{\uparrow}$ and $\Gamma^{\rm qp}_{\downarrow}$,  which satisfy
detailed balance condition $k_BT_S\approx (E_1-E_0)\ln(\Gamma^{\rm qp}_{\downarrow}/\Gamma^{\rm qp}_{\uparrow})$. 
Therefore, the levels with the decay rates $\Gamma_{n}<\Gamma^{\rm qp}_{\downarrow}$ become thermally populated
with the  temperature $T_S$, and the dark count rate grows. 
In order to estimate this effect, we ignore the anharmonicity of the potential well and assume that the level splittings and quasiparticle 
transition rates are the same for all levels. Afterwards, we obtain temperature dependent rate as
\begin{eqnarray}
\Gamma_{\rm dark}(T_S) \approx \Gamma_{\rm dark} + \Gamma^{\rm qp}_{\uparrow} W_{n_0}.
\label{Gamma_d1}
\end{eqnarray}
Here $n_0$ is the level number such that $\Gamma_{n_0-1}<\Gamma^{\rm qp}_{\downarrow}<\Gamma_{n_0}$, 
\begin{eqnarray}
W_{n0}\approx\frac{1-e^{-(E_1-E_0)/k_BT_S}}{e^{(E_{n_0}-E_0)/k_BT_S}-1}
\end{eqnarray}
is the thermal population of this level obtained assuming the normalization condition $\sum_{n=0}^{n_0}W_n=1$.
The latter condition follows from the observation that the level $n_0+1$ and the higher ones are not populated because of their fast decay.
The quasiparticle transition rates have been derived in Ref. \cite{Catelani} and read
\begin{eqnarray}
\Gamma^{\rm qp}_{\uparrow} &=& \frac{\Gamma_0^{\rm qp}}{e^{(E_1-E_0)/k_BT_S}-1},\;\;\;
\Gamma^{\rm qp}_{\downarrow} = \frac{\Gamma_0^{\rm qp}}{1-e^{-(E_1-E_0)/k_BT_S}},
\nonumber\\
\Gamma^{\rm qp}_0 &=& \frac{1}{e^2 R}\left|\left\langle 1 \left|\sin\frac{\varphi}{2}\right| 0 \right\rangle\right|^2
\sqrt{\frac{2\Delta}{E_1-E_0}} \frac{n_{\rm qp}}{\nu_0}.
\label{Gamma_qp0}
\end{eqnarray}
Here $n_{\rm qp}$ is the concentration of quasiparticles and $\nu_0$ is the density of states per unit spin in the 
superconductor. At low temperatures one finds $n_{\rm qp}=2\nu_0 \sqrt{2\pi\Delta k_BT_S}\,e^{-\Delta/k_BT_S}$. 
It has been found experimentally 
that in aluminum, for example, it is very difficult to reduce $T_S$ below $120$ \emph{mK} due to the presence of residual quasiparticles
with the lowest reported concentration $n_{qp}/(2\nu_0\Delta) \sim 10^{-9}$ \cite{Serniak,Elsa}.
Evaluating the matrix element of $\sin(\varphi/2)$, we transform Eq. (\ref{Gamma_qp0}) to
\begin{eqnarray}
\Gamma^{\rm qp}_{0} = \frac{\Delta e^{-\frac{\Delta}{k_BT_S}}}{e^2 R}
\left(\frac{128 \pi^2 k_B^2T_S^2 E_C}{E_J^3}\right)^{\frac{1}{4}}
\left(1+\frac{I_C}{\sqrt{I_C^2-I_b^2}}\right).
\nonumber\\
\label{Gamma_qp}
\end{eqnarray}

Next, we discuss the effect of unwanted thermal excitations in the resonator. For this purpose, 
we consider the time evolution of the occupation probabilities of the states $|10\rangle$ and $|0n_r\rangle$ with one and zero photons
in the resonator, which we denote as $p_1$ and $p_0$. 
For simplicity, we  ignore quantum coherence and describe the system by the two rate equations,
\begin{eqnarray}
\dot p_0 &=& -(\Gamma_{\rm dark} + \Gamma^{\rm qp}_{\uparrow} W_{n_0} + \kappa_r N_r  )p_0 + \kappa_r (N_r+1) p_1,
\nonumber\\
\dot p_1 &=& \kappa_r N_r\, p_0 - (\Gamma_{\rm sw} + \Gamma^{\rm qp}_{\downarrow} + \kappa_r (N_r+1)  )\,p_1.
\label{master}
\end{eqnarray}
Here  $N_r=(e^{\hbar\omega_r/k_BT_r}-1)^{-1}$ is the Bose function containing the resonator temperature,
$\kappa_r N_r$ is the rate of photon absorption by the resonator from its' dissipative environment
and $\kappa_r (N_r+1)$ is the rate of spontaneous photon emission to the environment. 
From Eqs. (\ref{master}) 
one finds that the state with zero photons decays as $p_0(t)\propto \exp(-\Gamma_{\rm dark}(T_r,T_S) t)$, where
\begin{eqnarray}
\Gamma_{\rm dark}(T_r,T_S) = \Gamma_{\rm dark} + \kappa_rN_r + \Gamma^{\rm qp}_{\uparrow} W_{n_0} 
\label{Gamma_full}
\end{eqnarray} 
is the total dark count rate estimated in the limit $\Gamma_{\rm dark}(T_r,T_S)\ll \Gamma_{\rm sw}$, and
the rate $\Gamma_{\rm dark}$ given by Eq. (\ref{Gamma_dark}).
Comparing the last two terms of Eq. (\ref{Gamma_full}) with the zero temperature dark count rate $\Gamma_{\rm dark}$,
we estimate the temperatures $T_r^*$ and $T_S^*$, below which the environment and the quasiparticle contributions can be ignored
and the detector should demonstrate its' best performance,
\begin{eqnarray}
&& T_r^* = \frac{\hbar\omega_r}{k_B\ln\left(1+\frac{\kappa_r}{\Gamma_{\rm dark}}\right)},
\label{Tr_star}
\\ &&
T_S^* \approx \frac{\Delta+(n_0+1)(E_1-E_0)}{k_B\ln\left[ \frac{\Delta}{e^2\Gamma_{\rm dark}}\sqrt{\frac{8\pi}{RR_q}}
\frac{E_C^{1/4}}{E_J^{1/4}}\left(1+\frac{I_C}{\sqrt{I_C^2-I_b^2}}\right) \right]}.
\label{TS_star}
\end{eqnarray}

Finally, we estimate the detector efficiency $\eta$, which
can be obtained from the second of the Eqs. (\ref{master}).
Indeed, according to it, after a photon absorption event the probability $p_1$
decays in time as $p_1(t)=e^{-(\Gamma_{\rm sw} + \Gamma^{\rm qp}_{\downarrow} + \kappa_r (N_T+1))t}$ until the photon
is dissipated by the environment, by quasiparticles or the junction switches to the resistive state. The probability
of the latter event determines the detector efficiency,
\begin{eqnarray}
\eta = \Gamma_{\rm sw}\int_0^\infty dt\,p_1(t) = \frac{\Gamma_{\rm sw}}{\Gamma_{\rm sw} + \Gamma^{\rm qp}_{\downarrow} + \kappa_r (N_T+1)}.
\label{eta}
\end{eqnarray}
As expected, the efficiency drops with the temperatures $T_r$, $T_S$ and with the damping rate of the resonator $\kappa_r$.

\section{Results of numerical simulation}
\label{Numerics}

In this section we present the results of numerical simulations described in
Sec. \ref{Numerics1} and compare them with simple approximations of Sec. \ref{Analytics}.
We consider  two sets of  system parameters. First, we choose the parameters typical for the circuit
quantum electrodynamics experiments. Afterwards, we consider the parameter values at which the detector
performace is significantly enhanced, but which may be more difficult to realize in experiment.

\begin{figure}
\includegraphics[width=\columnwidth]{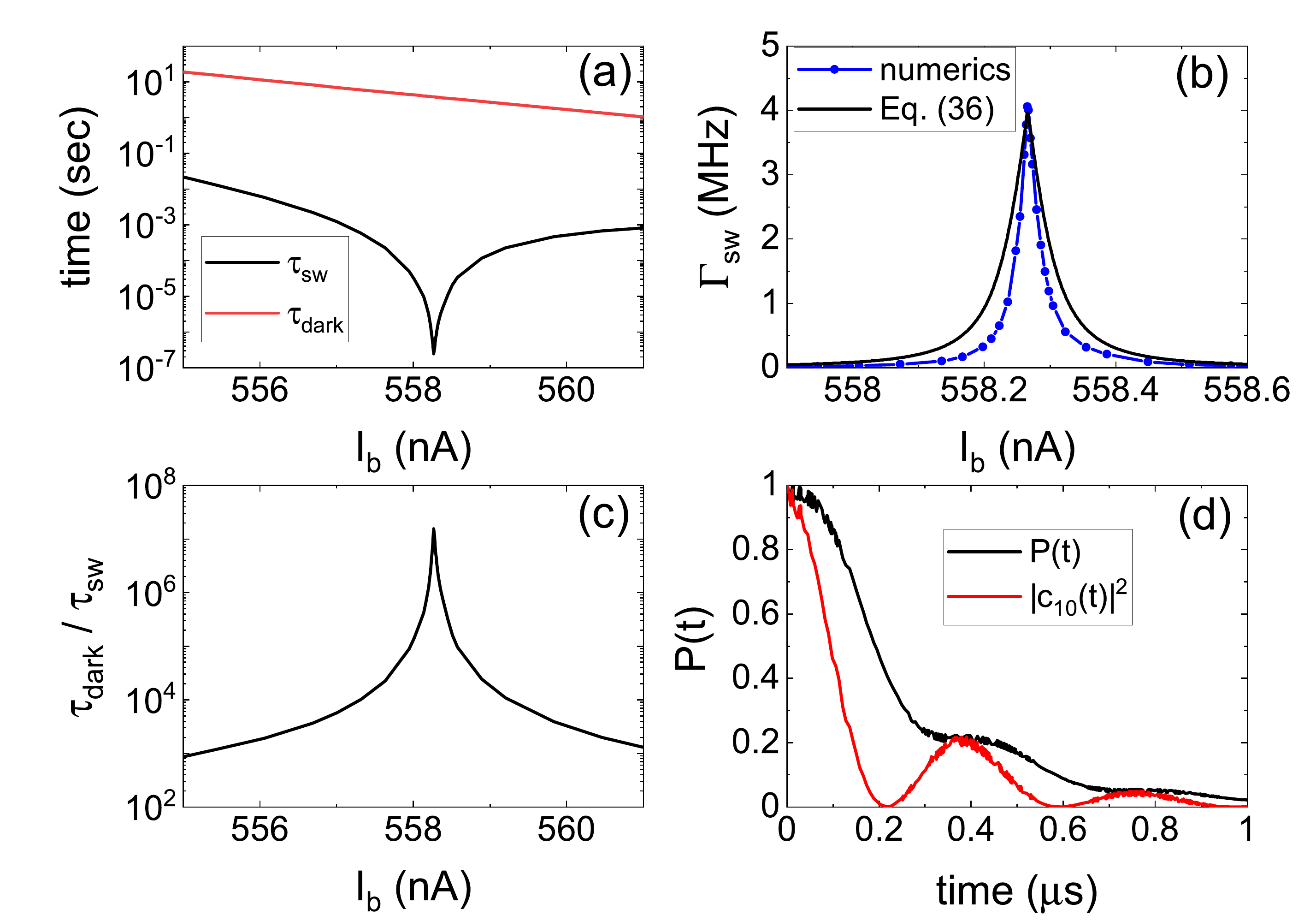}
\caption{(a) Dependence of the average time between dark counts ($\tau_{\rm dark}$, red line) and of the switching
time ($\tau_{\rm sw}$, black line) on the bias current $I_b$.
(b) Bias dependence of the switching rate $\Gamma_{\rm sw}=\tau_{\rm sw}^{-1}$ in the vicinity of the bias current
at which the resonance condition $\hbar\omega_r=\tilde E_3-\tilde E_0$ is met. The blue dots show the result of the numerical solution
and the black line -- the approximate analytical expression (\ref{Gamma_sw}) with the frequency converted to the current as in Eq. (\ref{I12}).
(c) The ratio $\tau_{\rm dark}/\tau_{\rm sw}$ versus bias current.
(d) Black line -- the time dependence of the probability $P(t)$, given by Eq.
(\ref{Pt}); red line -- the occupation probability $|c_{10}(t)|^2$ of the initial state $\psi_0=| 10 \rangle$
whith one photon in the resonator and ground state in the well of tilted Josephson potential.}
\label{plots}
\end{figure}

In Fig. \ref{plots}(a) we plot  $\tau_{\rm dark}$
and  $\tau_{\rm sw}$ as functions of the bias current $I_b$ for the first set of parameters.
In this simulation the system parameters are: the normal state resistance of the junction is $R_N=500$ $\Omega$,
the characteristic impedance of the transmission line resonator is $Z_0=50$ $\Omega$, 
the resonator frequency is $\omega_r/2\pi=14.5$ \emph{GHz}, 
the junction capacitance is $C=0.8$ \emph{pF},
and the coupling capacitance is $C_K=10$ \emph{fF}. Assuming that superconducting leads of the junction are made of aluminum with the gap value
$\Delta = 200$ \emph{$\mu$eV}, we find the critical current \emph{$I_C=\pi\Delta/2eR_N\cong 0.6$ $\mu$A }and the Josephson energy
$E_J/(2\pi\hbar)\cong 300$ \emph{GHz}. The charging energy of the junction takes the value $E_C/(2\pi\hbar) \cong 24.2$ \emph{MHz}, hence
we obtain the ratio \emph{$E_J/E_C=1.3\times 10^4$. }The McCumber paramters of such junction is big, \emph{$\beta=2eI_CR_N^2(C+C_K)/\hbar \cong 400$,}
which implies strongly hysteretic current-voltage characteristics favorable for single photon detection.
The coupling constant between the junction and the resonator (\ref{g})  is found to be ${g}/{2\pi} \cong 188.5$ \emph{MHz}.
At bias current $I_b\approx 558$ \emph{nA} the resonator frequency becomes equal to the transition frequency between the
levels 0 and 3, $\omega_r=(\tilde E_3-\tilde E_0)/\hbar$.
Thus in this run of the  numerical simluation we choose  $n_r=3$. There are 5 levels in the potential well, i.e. $N=4$,
in the whole interval of bias currents shown in Figs.  \ref{plots} (a,b) and (c).
At bias current corresponding to the resonance
the coupling constant for the  transition $0\leftrightarrow 3$, defined in Eq. (\ref{g_mn}),
takes the value \emph{$g_{03}/2\pi = 1.5$ MHz.}
We find the shortest switching time, achieved at resonace, to be $\tau_{\rm sw}^{\min} \cong 0.25$ $\mu$s, while the
dark count time in this case becomes \emph{$\tau_{\rm dark}=3.8$ s.}
The obtained value of the minimum switching time is comparable to dephasing times measured in good
quality phase qubits \cite{Martinis4} $T_2\approx 0.3$ $\mu$s, 
thus satisfying the condition $\tau_{\rm sw}^{\min}<T_2$, which is desireable for higher  detection efficiency.
We can estimate the quality factor of the resonator required for
the reliable photon detection as $Q_r>\omega_r\tau_{\rm sw}^{\rm min} \cong 2.2\times 10^4$. At lower quality
factor a photon is dissipated in the resonator earlier than the junction switches and
the detection efficiency (\ref{eta}) drops. For these parameters the dark count rate is unaffected by thermal population of the resonator
for temperatures $T_r\lesssim T_r^* \approx 42$ \emph{mK}, where $T_r^*$ is defined in Eq. (\ref{Tr_star}).

In Fig. \ref{plots}(b) we plot the swithing rate, $\Gamma_{\rm sw}=\tau_{\rm sw}^{-1}$, in the vicinity of the resonance with blue dots.
For comparison, we also plot the analytical formula (\ref{Gamma_sw}) with the black line. Both curves have
the shape of a peak with a cusp. We have used the following input parameters for the analytical model:
the escape rate from the third level of an uncoupled junction (\ref{Gamma}),
$\Gamma_3=7.2$ \emph{MHz}, the same escape rate enhanced by the coupling to the gound state of the
resonator (\ref{Gamma_nr}), $\tilde \Gamma_{03}=8$ \emph{MHz},
and the escape rate for the state $|10\rangle$ (\ref{Gamma_0}), $\tilde\Gamma_{10}=1.6$ \emph{Hz}.
Since $g_{03}>\tilde \Gamma_{03}/4$ and $\tilde \Gamma_{03} \gg \tilde\Gamma_{10}$,
the maximum switching rate (\ref{Gamma_max}) is expected to be
$\Gamma_{\rm sw}^{\max}\approx\tilde\Gamma_{03}/2=4$ \emph{MHz}, which perfectly agrees with the numerics.
The half width of the peak in the frequency units (\ref{D12}) takes the value $\delta\omega_{1/2}=21$ \emph{MHz}.
It translates to the width of the current peak (\ref{I12}) $\Delta I_{b,1/2}=63$ \emph{pA}, which is
approximately two times bigger than the value obtained numerically,  $\Delta I_{b,1/2}=35$ \emph{pA}.
The descrepancy between the analytical model and the numerics comes from rather simple approximation
for the level splitting $\tilde E_{n_r}-\tilde E_0\approx n_r\sqrt{8E_JE_C\alpha}$, which we  used to
derive the frequency to current conversion factor in Eq. (\ref{I12}). This approximation, however, is sufficiently
accurate for a rough estimate of the current peak width.
In Fig. \ref{plots}(c) we plot the ratio $\tau_{\rm dark}/\tau_{\rm sw}$, which characterizes
the performance of the detector, as a function of the bias current. The maximum ratio is achieved at resonance,
$\tau_{\rm dark}/\tau_{\rm sw}=1.54\times 10^7$. It is slightly higher than  the estimate (\ref{ratio}), 
which predicts $\tau_{\rm dark}/\tau_{\rm sw}=9\times 10^6$. 

So far we have ignored the effect of quasiparticles in superconducting leads. 
Assuming the lowest   
effective temperature of the aluminum leads of the junction reported in qubit experiments, $T_S=120$ \emph{mK}, from Eqs. 
(\ref{Gamma_qp0},\ref{Gamma_qp})
we obtain $\Gamma_{\downarrow}^{\rm qp}=1.8$ \emph{kHz}. Next, we numerically find the decay rates of the first and the second level in the well,
$\Gamma_{1}\approx 35$ \emph{Hz}, $\Gamma_{2}\approx 22.6$ \emph{kHz} and observe that $\Gamma_1<\Gamma_{10}^{\rm qp}<\Gamma_2$.
Thus, we set $n_0=2$ in Eq. (\ref{Gamma_full}), which results in the dark count rate 
$\Gamma_{\rm dark}(0,T_S)=3$ \emph{Hz}, dark count time $\tau_{\rm dark}=0.36$ \emph{s} and the ratio 
$\tau_{\rm dark}/\tau_{\rm sw}\approx 1.4\times 10^6$. Thus, the presence of non-equilibrium quasiparticles slightly reduces the dark count time
of the device. We find that one should cool the junction leads
below the temperature (\ref{TS_star}) $T_S^*=110$ \emph{mK} in order to achieve the dark count time 3.8 \emph{s} reported above. 
Such temperatures may be easier to achieve in the proposed detector than in qubit devices 
because the leads of the junction, which are not electrically isolated superconducting islands, can be made sufficiently bulky. 
In addition, in our setup one can use normal metal quasiparticle traps in more straightforward manner. 

In Fig. \ref{plots}(d) we plot the time dependence
of the probability $P(t)$ (\ref{Pt}) at resonance (black line). We also show the occupation probability
of the initial state $|10\rangle$, in which the resonator hosts one photon and the junction is in its ground state,
which we denote as $|c_{10}(t)|^2$. This probability oscillates because of the coherent 
coupling between the resonator and the junction. The frequency of these oscillations equals to ${\rm Re}(\Omega)$,
where $\Omega$ is defined in Eq. (\ref{Omega}). For the system parameters given above we find ${\rm Re}(\Omega)/2\pi = 1.42$ \emph{MHz}.
Both functions shown in Fig. \ref{plots}(d) also exhibit high frequency small amplitude oscillations, 
which are not captured by the analytical expression (\ref{Pt2}).

\begin{figure}
\includegraphics[width=\columnwidth]{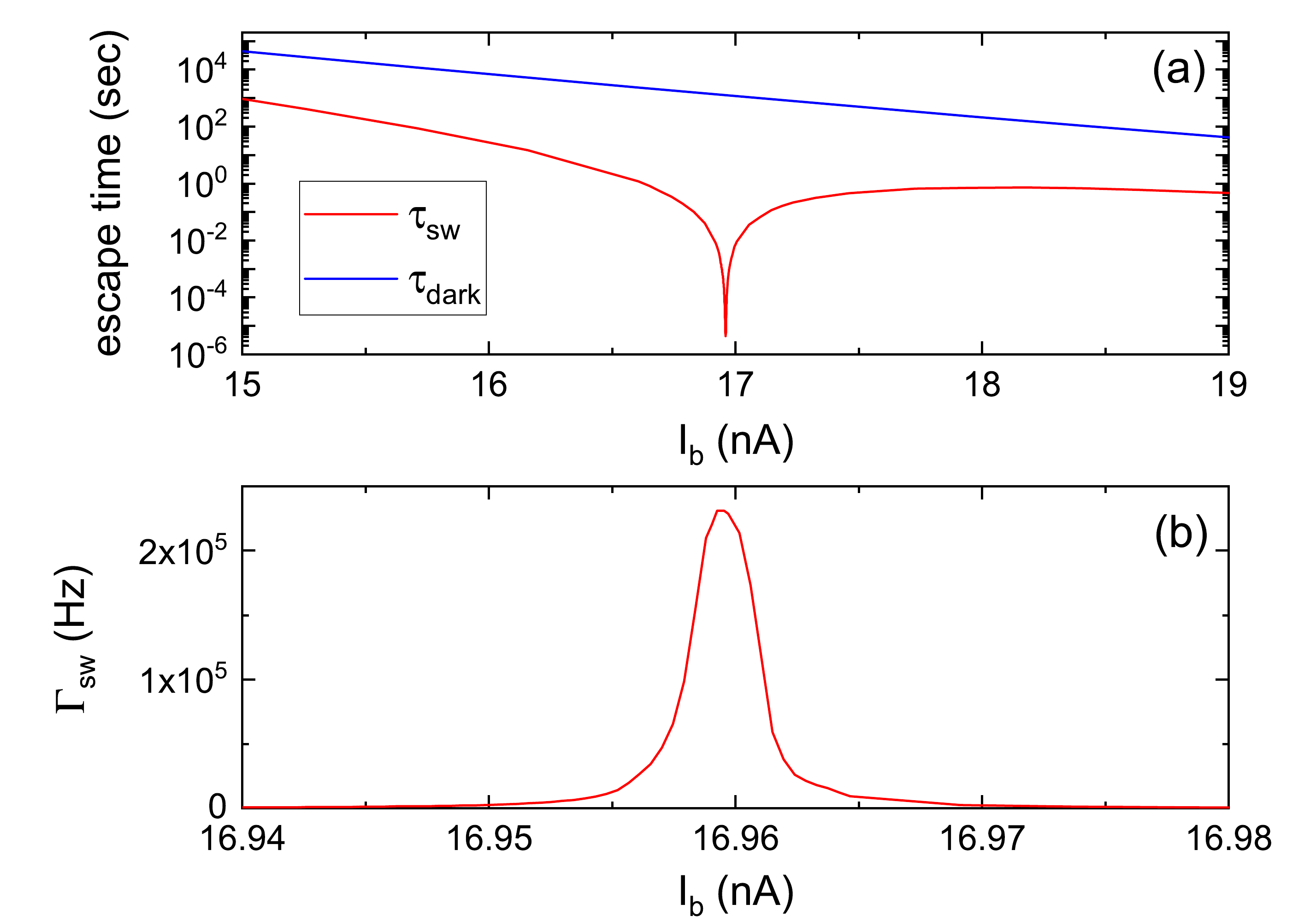}
\caption{(a) Dependence of the average time between dark counts ($\tau_{\rm dark}$, blue line) and of the switching
time ($\tau_{\rm sw}$, red line) on the bias current $I_b$.
(b) Dependence of the switching rate $\Gamma_{\rm sw}=\tau_{\rm sw}^{-1}$ on the bias current in the vicinity
of $I_b=16.96$ nA, at which the resonance condition $\hbar\omega_r=\tilde E_4-\tilde E_0$ occurs. }
\label{Fig2}
\end{figure}

Next, we consider another set of paramters and assume that the level number 4 is alinged with the resonator, 
i.e. we choose $n_r=4$. We also choose $N=5$, which means there are 6 levels in the well.
The parameters of the system are: $R_N=7$ k$\Omega$,
 $Z_0=50$ $\Omega$, $\omega_r/2\pi=14.5$ \emph{GHz}, $C=200$ \emph{fF}
and $C_K=5$ \emph{fF}. This results in the critical current $I_C\cong 45$ \emph{nA} and the Josephson energy
$E_J/(2\pi\hbar)\cong 22$ \emph{GHz}. The charging energy of the junction takes the value $E_C/(2\pi\hbar) \cong 97$ \emph{MHz}, hence
we obtain the ratio $E_J/E_C=230$. The McCumber paramters of such a junction is very high, $\beta=2eI_CR_N^2(C+C_K)/\hbar \cong 1300$.
With these parameters,  the coupling constant (\ref{g}) between the junction and the resonator takes the value
${g}/{2\pi} \cong 260$ \emph{MHz}.
At bias current $I_b=16.96$ \emph{nA} the resonance condition  $\hbar\omega_r=\tilde E_4-\tilde E_0$ is achieved.
At this bias point
the coupling constant for the  transition $0\leftrightarrow 4$, given by Eq. (\ref{g_mn}), is rather small, $g_{04}/2\pi = 84$ \emph{kHz},
which makes the experimental observation of the $0\leftrightarrow 4$ transitions difficult.
In Fig. \ref{Fig2}(a) we plot the times  $\tau_{\rm dark}$ and $\tau_{\rm sw}$
as functions of the bias current $I_b$ for this set of parameters.
We find the shortest switching time, achieved at resonace, to be $\tau_{\rm sw}^{\min} \cong 4$ $\mu$s and the
dark count time -- $\tau_{\rm dark}=21$ \emph{min}. The ratio between these two times 
is very large, $\tau_{\rm dark}/\tau_{\rm sw}=7.5\times 10^8$. The estimate (\ref{ratio})
for this case predicts the ratio $4\times 10^9$. As expected, by choosing $n_r=4$ we have significantly increased the
ratio $\tau_{\rm dark}/\tau_{\rm sw}$ as compared to the previous set of parameters with $n_r=3$.
The quality factor of the resonator required for
the reliable operation of the detector is $Q_r>\omega_r\tau_{\rm sw}^{\rm min} \cong 4\times 10^5$.
With this  quality factor the upper bound for the temperature (\ref{Tr_star}) is $T_r\lesssim 30$ \emph{mK}.

Let us now discuss the effect of quasiparticles.
Assuming again the effective temperature of the aluminum leads of the junction $T_S=120$ \emph{mK}, 
from Eqs. (\ref{Gamma_qp0},\ref{Gamma_qp})
we find $\Gamma_{\downarrow}^{\rm qp}=950$ \emph{Hz}. Numerically we find 
$\Gamma_{2}\approx 7$ \emph{Hz} and $\Gamma_{3}\approx 3.7$ \emph{kHz}, which means $\Gamma_2<\Gamma_{10}^{\rm qp}<\Gamma_3$.
Hence, we  put $n_0=3$ in Eq. (\ref{Gamma_full}) and obtain the dark count rate 
$\Gamma_{\rm dark}(T)=1.2$ \emph{Hz}, dark count time $\tau_{\rm dark}=0.8$ \emph{s} and the ratio 
$\tau_{\rm dark}/\tau_{\rm sw}\approx 2\times 10^5$. Thus, for this set of parameters the residual
quasiparticles in the leads of the junction significantly degrade the performance of the detector.
We find that in oder to approach zero temperature value of $\tau_{\rm dark}$  the junction leads
should be cooled below  $T_S^*=93$ \emph{mK}, which follows from Eq. (\ref{TS_star}).

In Fig. \ref{Fig2}(b) we plot the switching rate  in the vicinity of the resonance.
It has the form of a narrow peak with the maximum height $\Gamma_{\rm sw}^{\max}\approx 230$ \emph{kHz}
and the half-width $I_{b,1/2}\approx 2.2$ \emph{pA}.
The analytical model (\ref{Gamma_sw},\ref{D12},\ref{I12})
with the input parameters $\tilde\Gamma_{04}=1.1$ \emph{MHz} and $\tilde\Gamma_{10}=1.3$ \emph{mHz}
predicts $\Gamma_{\rm sw}^{\max}\approx 575$ \emph{kHz},
$\delta\omega_{1/2}\approx 1.1$ \emph{MHz}, and $\Delta I_{b,1/2}\approx 2.2$ \emph{pA}.
In this case, the approximate model overstimates the maximum switching rate due to the slower non-exponential decay
of the probability $P(t)$ at short times. On the other hand, in this case the analytical model very accurately predicts the width of the peak.
This width turns out to be small, which makes practical realization of the detector with these parameters difficult.

The two examples considered above illustrate that one can push the ratio $\tau_{\rm dark}/\tau_{\rm sw}$ to
very high values by increasing the number of the resonant level $n_r$.
However, by doing so one simultaneously decreases the width of the resonance peak in current units. One can partly
compensate for that by choosing larger coupling capacitor $C_K$ and in this way increasing the coupling constant $g$.

\section{Conclusion}
\label{Summary}

We have proposed and theoretically analyzed a single photon detector in the microwave frequency range,
which consists of a current biased Josephson junction coupled to a high quality factor resonator.
We have shown that for typical system parameters the ratio between the switching rate after a photon arrival
and the dark count rate can achieve the value $\Gamma_{\rm sw}/\Gamma_{\rm dark}\sim 10^7$
provided the superconducting leads of the Josephson junction are cooled below 90 \emph{mK} and
the environment of the resonator -- below 30 \emph{mK}. With some effort,
it should be possible to achieve even higher ratios  $\Gamma_{\rm sw}/\Gamma_{\rm dark}\sim 10^9$.
Such a detector can operate at very low photon fluxes, where the time intervals between the photons may reach
seconds or even hours. It can be useful in the detection of rare events like, for example, decay of elementary particles.

This work was supported by the Academy of Finland Centre of Excellence program (Project No. 312057)
and by the European Union's Horizon 2020 research and innovation programme under grant agreement No. 863313 (SUPERGALAX).
It was also partly supported by the Ministry of Science and Higher Education of the Russian Federation (Grant No. FSUN-2020-0007)
and by the Russian Science Foundation (Project No. 19-79-10170).

\appendix

\section{Derivation of Eq. (\ref{H})}
\label{App_H}

In this section we derive the Hamiltonian (\ref{H}).
We follow the standard procedure, which has been used, for example, in Refs. \cite{Andresen2,Bourassa} for slightly different systems.
The classical equations (\ref{sys}), with dissipative terms omitted, can be derived from the Lagrangian
\begin{eqnarray}
{\cal L} &=& \frac{C+C_K}{2}\left(\frac{\hbar\dot\varphi}{2e}\right)^2 - E_J(1-\cos\varphi) + \frac{\hbar I_b}{2e}\varphi
\nonumber\\ &&
+\, \frac{\pi(\dot V_K^2 - \omega_r^2 V_K^2)}{4Z_0\omega_r^3} + C_K\frac{\hbar\varphi}{2e}\dot V_K.
\label{L}
\end{eqnarray}
We find the corresponding classical momenta
\begin{eqnarray}
p_\varphi &=& \frac{\partial{\cal L}}{\partial\dot\varphi} = (C+C_K)\frac{\hbar^2\dot\varphi}{4e^2},
\nonumber\\
p_{V_k} &=& \frac{\partial{\cal L}}{\partial\dot V_K} = \frac{\pi \dot V_K}{2Z_0\omega_r^3} + C_K\frac{\hbar\varphi}{2e},
\end{eqnarray}
and the classical Hamiltonian
\begin{eqnarray}
H &=& p_{\varphi}\dot\varphi + p_{V_K}\dot V_K - {\cal L}
\nonumber\\
&=&  \frac{4e^2}{\hbar^2}\frac{p_\varphi^2}{2(C+C_K)} + E_J(1-\cos\varphi) - \frac{\hbar I_b}{2e}\varphi
\nonumber\\ &&
+\, \frac{Z_0\omega_r^3}{\pi}\left(p_{V_k} - C_K\frac{\hbar\varphi}{2e}\right)^2 + \frac{\pi V_K^2}{4Z_0\omega_r}.
\label{H_cl}
\end{eqnarray}

The quantum Hamiltonian (\ref{H}) is obtained from the classical one (\ref{H_cl}) by making the replacements
\begin{eqnarray}
&& p_\varphi \to  -i\hbar\frac{\partial}{\partial\varphi}, \;\;
p_{V_K} \to i\sqrt{\frac{\pi\hbar\omega_r}{4Z_0\omega_r^3}}(\hat a^\dagger-\hat a),
\nonumber\\
&& V_K \to \sqrt{\frac{\hbar\omega_r^2Z_0}{\pi}}(\hat a^\dagger+\hat a),
\end{eqnarray}
where $\hat a^\dagger$ and $\hat a$ are the creation and annihilation operators of the photons in the resonator.

\end{document}